\documentclass[12pt,titlepage]{article}         

\usepackage[american]{babel}
\usepackage[T1]{fontenc}
\usepackage{times}
\usepackage[dvips,usenames]{color}
\usepackage[dvips]{graphics}
\usepackage{graphicx}
\usepackage{subfigure}
\usepackage{amsmath}
\usepackage{amssymb}
\usepackage{natbib}
\bibpunct[, ]{(}{)}{;}{a}{}{,}
\usepackage{ifthen}

\newboolean{includefigs}
\setboolean{includefigs}{true}
\newcommand{\condcomment}[2]{\ifthenelse{#1}{#2}{}}

\usepackage[dvips,usenames]{color}
\definecolor{DarkBlue}{rgb}{0.0,0.0,0.5}
\definecolor{DarkRed}{rgb}{0.5,0.0,0.0}
\definecolor{DarkGreen}{rgb}{0.0,0.5,0.0}

\def\bOmega{{\mathbf{\Omega}}}
\def\jj{{\mathbf{j}}}
\def\bu{{\mathbf{U}}}
\def\bn{{\mathbf{\nabla}}}

\def\blu{{\mathbf{u}}}

\def\bb{{\mathbf{B}}}
\def\obb{{\overline{\mathbf B}}}
\def\obu{{\overline{\mathbf U}}}
\def\out{{{\overline U}_\phi}}
\def\op{{\overline p}}
\def\orho{{\overline\rho}}
\def\oPhi{{\overline\Phi}}
\def\oba{{\overline{\mathbf a}}}
\def\obg{{\overline{\mathbf g}}}
\def\os{{\overline s}}
\def\bv{{\mathbf v}}

\def\blb{{\mathbf{b}}}

\def\br{{\mathbf{r}}}

\def\bj{{\mathbf{j}}}
\def\be{{\mathbf{E}}}
\def\bff{{\mathbf{F}}}
\def\bn{{\mathbf{\nabla}}}

\newcommand{\m}{\ \rm{m}}
\newcommand{\cm}{\ \rm{cm}}
\newcommand{\s}{\ \rm{s}}

\newcommand{\K}{\ \rm{K}}
\newcommand{\bigS}{\ \rm{S}}

\newcommand{\g}{\ \rm{g}}

\newcommand{\mathbar}{\ \rm{bar}}
\newcommand{\Mbar}{\ \rm{Mbar}}

\newcommand{\beq}{\begin{equation}}
\newcommand{\eeq}{\end{equation}}
\newcommand{\beqa}{\begin{eqnarray}}
\newcommand{\eeqa}{\end{eqnarray}}
\newcommand{\de}{\partial}

\newcommand{\ep}{\mathbf{e}_{\phi}}
\newcommand{\ez}{\mathbf{e}_{z}}

\newboolean{preprint}
\setboolean{preprint}{false}

\ifthenelse{\boolean{preprint}}{
\usepackage[preprint]{amsstyle}
\pagestyle{myheadings}
\markboth{J.\ J. Liu, P.\ M Goldreich and D.\ J. Stevenson}{Ohmic Dissipation Constraint}
}{ 
\usepackage[draft]{amsstyle}
\usepackage[nomarkers]{endfloat}
\AtBeginDelayedFloats{}   
\AtBeginFigures{}         
} 

\reference{To be submitted to \textit{Icarus}}

\begin{document}


\begin{titlepage}
  \begin{center}
   { \Large{Constraints on Deep-seated Zonal Winds Inside
Jupiter and Saturn }}

\bigskip

Junjun Liu

{\it 100-23, Geological and Planetary Sciences, Caltech, Pasadena, CA 91125. \\ E-mail: ljj@gps.caltech.edu}

\bigskip

Peter M. Goldreich

{\it  School of Natural Sciences, Institute for Advanced Study, Princeton, NJ 08540.  \\ E-mail: pmg@ias.edu}

\bigskip

David J. Stevenson

{\it 150-21, Geological and Planetary Sciences, Caltech, Pasadena, CA 91125. \\ E-mail: djs@gps.caltech.edu}

\bigskip \bigskip \bigskip

{\bf Submitted to Icarus.}

 \end{center}

Manuscript pages including references and figure captions: 24. Number of figures: 12.

Keywords: (1) MAGNETIC FIELDS, (2) ATMOSPHERES, DYNAMICS, (3) INTERIORS.
\end{titlepage}

Proposed running head:

Constraints on Deep-seated Zonal Winds Inside Jupiter and Saturn

\bigskip \bigskip \bigskip \bigskip
Send proofs and correspondence to:

Junjun Liu

100-23, Geological and Planetary Sciences,

Caltech,

Pasadena, CA 91125.

E-mail: ljj@gps.caltech.edu

Phone: (626)-395-6894

Fax: (626)-585-1917

\newpage

\begin{abstract}
The atmospheres of Jupiter and Saturn exhibit strong and stable
zonal winds. How deep the winds penetrate unabated into each planet
is unknown. Our investigation favors shallow winds. It consists of
two parts.

The first part makes use of an Ohmic constraint; Ohmic dissipation
associated with the planet's magnetic field cannot exceed the
planet's net luminosity. Application to Jupiter (J) and Saturn (S)
shows that the observed zonal winds cannot penetrate below a depth
at which the electrical conductivity is about six orders of
magnitude smaller than its value at the molecular-metallic
transition. Measured values of the electrical conductivity of
molecular hydrogen yield radii of maximum penetration of $0.96R_J$
and $0.86R_S$, with uncertainties of a few percent of $R$. At these
radii, the magnetic Reynolds number based on the zonal wind velocity
and the scale height of the magnetic diffusivity is of order unity.
These limits are insensitive to difficulties in modeling turbulent
convection. They permit complete penetration along cylinders of the
equatorial jets observed in the atmospheres of Jupiter and Saturn.

The second part investigates how deep the observed zonal winds
actually do penetrate. As it applies heuristic models of turbulent
convection, its conclusions must be regarded as tentative.
Truncation of the winds in the planet's convective envelope would
involve breaking the Taylor-Proudman constraint on cylindrical flow.
This would require a suitable nonpotential acceleration which none
of the obvious candidates appears able to provide. Accelerations
arising from entropy gradients, magnetic stresses, and Reynolds
stresses appear to be much too weak.  These considerations suggest
that strong zonal winds are confined to shallow, stably stratified
layers, with equatorial jets being the possible exception.

\end{abstract}

\newpage

\setcounter{page}{1}

\section{Introduction}

\label{intro}

Jupiter and Saturn, are composed primarily of hydrogen and helium
with small additions of heavier elements. Their atmospheres exhibit
strong, stable zonal winds composed of multiple jets associated with
azimuthal cloud bands \citep{ingersoll90}. Zonal winds peak in the
equatorial region reaching $\sim 100 \m \s^{-1}$ on Jupiter and
$\sim 400 \m \s^{-1}$ on Saturn.\footnote{Wind speeds on Jupiter are
determined relative to System III coordinates which rotate with the
angular speed of the planet's magnetic field \citep{dessler83}.
Only differences among wind speeds on Saturn are known because the planet's
internal rotaton rate is uncertain.} The
latitudes of Jupiter's jets have not changed for at least 80 years
\citep{smith76} and their velocities have been constant within
$10\%$ over $25$ years \citep{porco03}.  \\

The depth of the zonal winds is unknown. Both deep and shallow flow
models have been proposed.  Wind speeds measured by the Galileo
probe at $7.4^{\circ}N$ on Jupiter rose from $90\m\s^{-1}$ at $0.4
\mathbar$ to $180\m \s^{-1}$ at $\sim 5 \mathbar$ and then remained
nearly constant until $22 \mathbar$ \citep{atkinson97,atkinson98}.
It is important to bear in mind that these measurements only sample
the winds in the outer $1 \%$ of the planet's radius where the
electrical conductivity is low. In regions of high electrical
conductivity, the magnetic field lines are frozen into the fluid.
Winds in these regions would cause changes in the external magnetic
field. By comparing Galileo and Pioneer/Voyager data,
\citet{russell01a,russell01b} find that increases of $0.3\deg$ in
the dipole tilt and $1.5\%$ in Jupiter's dipole moment may have
taken place between 1975 and 2000. The former could be accounted for
by meridional flows with speeds in the deep interior of order
$0.1\cm\s^{-1}$ \citep{guillot04}.
\\

\citet{busse76,busse83,busse94} advocates deep flows. He applies the
Taylor-Proudman theorem \citep{taylor23} to deduce that zonal flows
extend along cylinders oriented parallel the rotation axis in the
molecular envelope, and then terminate at the outer boundary of the
metallic core where he assumes that hydrogen undergoes a first order
phase transition. But data from shock wave experiments show that
hydrogen undergoes a continuous transition from a semi-conducting
molecular state to a highly conducting metallic state as the
pressure increases. This contradicts the assumption of a first order
phase transition at the core-envelope boundary.  \\

Recently, a modified deep flow model for Jovian zonal flows has been
proposed based on simulations of convection in a shell with a lower
boundary near $0.9R_J$ \citep{aurnou04,heimpel05}. The physical
meaning of the lower boundary in the modified deep flow model is
obscure. Hydrogen cannot undergo a phase change at that radius
\citep{guillot04}. So how might the Taylor-Proudman constraint be
violated in order to reduce the zonal flow to a near zero value
below that boundary? We address related issues in \S
\ref{sec:truncate}.
\\

In shallow flow models, the observed high-speed flow is confined to
a thin, baroclinic layer near the cloud level; the interior flow is
much slower. Even if the high velocity flow is confined to a shallow
layer, its forcing may occur at depth. For example, if the flow were
to arise from a process that conserved angular momentum per unit
volume, $\rho U$ would be approximately conserved, where $\rho$ is
the density and $U$ is the magnitude of the flow velocity. Since the
density in the interior is several orders of magnitudes larger than
that near the surface, the flow velocity could then be much greater
near the surface. On the other hand, the observed zonal flow might
be generated by shallow forcing due to the turbulence injected at
the cloud level by moist convection, differential latitudinal solar
heating, latent heat release from condensation of water, or other
weather layer processes \citep{vasavada05}. From the
thermal wind equation, a latitudinal temperature gradient of about
$5-10K$ across a few pressure scale heights below the cloud level
would cause substantial vertical shear, which makes the flow
velocity much greater near the surface than deeper down
\citep{ingersoll69,ingersoll84,vasavada05}.
\\

The plan of our paper is as follows. Relevant details of the
electrical conductivity of molecular hydrogen as measured in shock
wave experiments are presented in \S \ref{odsec:conductivity}.
Sections \ref{sec:extrapolation} and \ref{sec:cylindrical} are
devoted to the calculation of Ohmic dissipation based on the
assumption that the zonal wind penetrates the planet along
cylinders. In the former, the poloidal magnetic field is determined
by downward extrapolation of the external field. This procedure is
appropriate in regions where the magnetic Reynolds based on the
convective velocity, $R_m^c\ll 1$. In the latter, we examine the
consequences of assuming that the poloidal magnetic field is
parallel to the rotation axis in regions where $R_m^c\gg 1$. The
requirement that the total Ohmic dissipation be bounded from above
by the planet's net luminosity, $\cal L$, limits the depth to which
the observed zonal winds can penetrate. Section \ref{sec:truncate}
asks whether the zonal winds might be truncated within the
convective envelope. A short summary of our main results is given in
\S \ref{sec:discussion}. A few technical details are relegated to
the Appendix.
\\

\section{Electrical Conductivity In Jupiter \& Saturn}
\label{odsec:conductivity}

Electrical conductivity in the interiors of Jupiter and Saturn is
due mainly to hydrogen. Near their surfaces it might be
significantly enhanced relative to pure hydrogen by the addition of
some more readily ionized heavier elements. Helium is unimportant
due to its high ionization potential.  \\

Condensed molecular hydrogen is a wide band-gap insulator at room
temperature and pressure, with a band gap, $E_g$, of about $15 \
\rm{eV}$, corresponding to the ionization energy of the hydrogen
molecule. As the pressure increases, this gap is expected to diminish
and finally close to zero, resulting in an insulator-to-metal
transition. In experiments, this transition appears to be
gradual. As the energy gap closes, hydrogen molecules begin to
dissociate to monatomic hydrogen and electrons start to be delocalized
from $\rm{H}_2^{+}$ ions \citep{nellis96,weir96}.
The insulator-to-metal transition is expected to occur even
though the hydrogen molecules have not been fully
pressure-dissociated. At much higher pressure and temperature,
molecular dissociation becomes complete and it is presumed that pure
monatomic hydrogen forms a metallic Coulomb plasma
\citep{stevenson74,hubbard97}.
\\

The conductivity of hydrogen has been measured in reverberating
shock wave experiments from $0.93 \Mbar$ to $1.8 \Mbar$
\citep{weir96,nellis99} and in single shock experiments
from $0.1 \Mbar$ to $0.2 \Mbar$ \citep{nellis92}. In these
experiments, hydrogen is in thermal equilibrium at pressures and
temperatures similar to those in the interiors of giant planets.
From $0.93 $ to $1.4$ Mbar, the measured electrical conductivity of
hydrogen increases by four orders of magnitude. Between $1.4 \Mbar$
and $1.8 \Mbar$, the conductivity is constant at $2 \times 10^5
\bigS \m^{-1}$, similar to that of liquid Cs and Rb at $2000\K$ and
two orders of magnitude lower than that of a good metal (e.g. Cu) at
room temperature. The constant conductivity suggests that the energy
gap has been thermally smeared out \citep{weir96}.
Temperatures of shock-compressed liquid hydrogen have been measured
optically in separate experiments \citep{nellis95,holmes95}.
At the highest obtained pressure of $0.83 \ \rm{Mbar}$,
the measured temperature of $5200 \K$ falls below that predicted for
pure molecular hydrogen. This is due to the dissociation of
molecular hydrogen and enables us to estimate the fractional
dissociation as a function of pressure.  At $1.4 \Mbar$ and $3000
\K$, the dissociation fraction is $ \sim 5\%$. Thus metallization of
hydrogen occurs in the diatomic molecular phase and is caused by
electrons delocalized from $\rm{H}_2^{+}$ ions \citep{nellis96,ashcroft68}.
Since we are interested in the outer shell of
the giant planets, the measurements at $0.1-0.2 \Mbar$ are the most
relevant. In this low-pressure range,
the dissociation of hydrogen molecules is unimportant. \\

The electrical conductivity of a semiconductor can be expressed in the
form:
\begin{equation}
\sigma = \sigma_0(\rho) \exp \left( - \frac{E_g(\rho)}{2 K_B T} \right),
\label{coneq: conductivity}
\end{equation}
where $\sigma$ is electrical conductivity, $E_g(\rho)$ is the energy
of the density dependent mobility gap, $K_B$ is Boltzmann's
constant, $T$ is the temperature, and $\exp \left( - E_g/2 K_B
T\right)$ expresses the fractional occupancy of the current carrying
states. The conductivity measurement at $0.27 \g \cm^{-3}$ and a
temperature of $4160 \K$ \footnote{The corresponding pressure is
$0.187 \Mbar$.} \citep{nellis92} is close to the interior isentropes
of Jupiter and Saturn \citep{guillot99}. This measurement determines
$\sigma_0 = 1.1 \times 10^8 \bigS \m^{-1}$ and $E_g = 11.7 \pm 1.1
\mathrm{eV}$. The error bar in the energy gap comes from the
experimental uncertainties in $\sigma$ and uncertainties in
calculated post-shock temperatures \citep{nellis95}. \\

The data suggest that it is sufficiently accurate to assume $E_g$ is
a linear function of hydrogen density $\rho_{H_2}$: $E_g = a + b
\rho_{H_2}$, where $a$ and $b$ are two constants
\citep{nellis92,nellis96}. The measurements and the energy gap ($15
\mathrm{eV}$) at room temperature and pressure determine the values
of $a$ and $b$. Since the volume mixing ratio of hydrogen in the
outer shells of giant planets is about $92 \%$, the density in giant
planet interiors is $1.18 \rho_{H_2}$. With $\sigma_0$,
$E_g(\rho_{H_2})$, and $T(\rho)$ from the planetary interior model
\citep{guillot99}, the conductivity profiles of the giant
planets can be calculated. \\

Figure (\ref{config: conductivity}) displays the conductivity
distributions as a function of radius in the outer shells of Jupiter
and Saturn. The conductivity increases with depth, and there is a
smooth transition from semi-conducting to metallic hydrogen at
pressure ($1.4 \ \mathrm{Mbar}$). This transition takes place at
about $0.84$ of Jupiter's radius and $0.63$ of Saturn's radius.
\\

The electrical conductivity is proportional to the total number
density of electrical charge carriers: $\sigma \propto n_e$, which
includes a contribution, which we have neglected, from impurities
$x$ in addition to that from hydrogen:
\begin{equation}
n_e = n_{H_2} \exp\left( - \frac{E_g}{2 K_B T} \right)
+ \sum_x n_x \exp \left( - \frac{E_x}{2K_BT} \right),
\end{equation}
where $n_x$ and $E_x$ express the number density of the electrons and
the energy gap due to an impurity. Alkali metals are sources of small
band gap impurities. They may also contribute to the radiative opacity
thus insuring adiabaticity \citep{guillot04,guillot05}. The
mixing ratio of an alkali metal in the interior of a giant planet is
presumably similar to that determined from its cosmic abundance. With
these abundances, a band gap of a few electron volts would lead to a
conductivity of $10^{-6} \sim 10^{-4}\bigS \m^{-1}$ at $T \sim 1000
\K$, significantly above the value due to hydrogen in the outer shells
of giant planets. \\

\begin{figure}
\centerline{\includegraphics[width=5.2in]{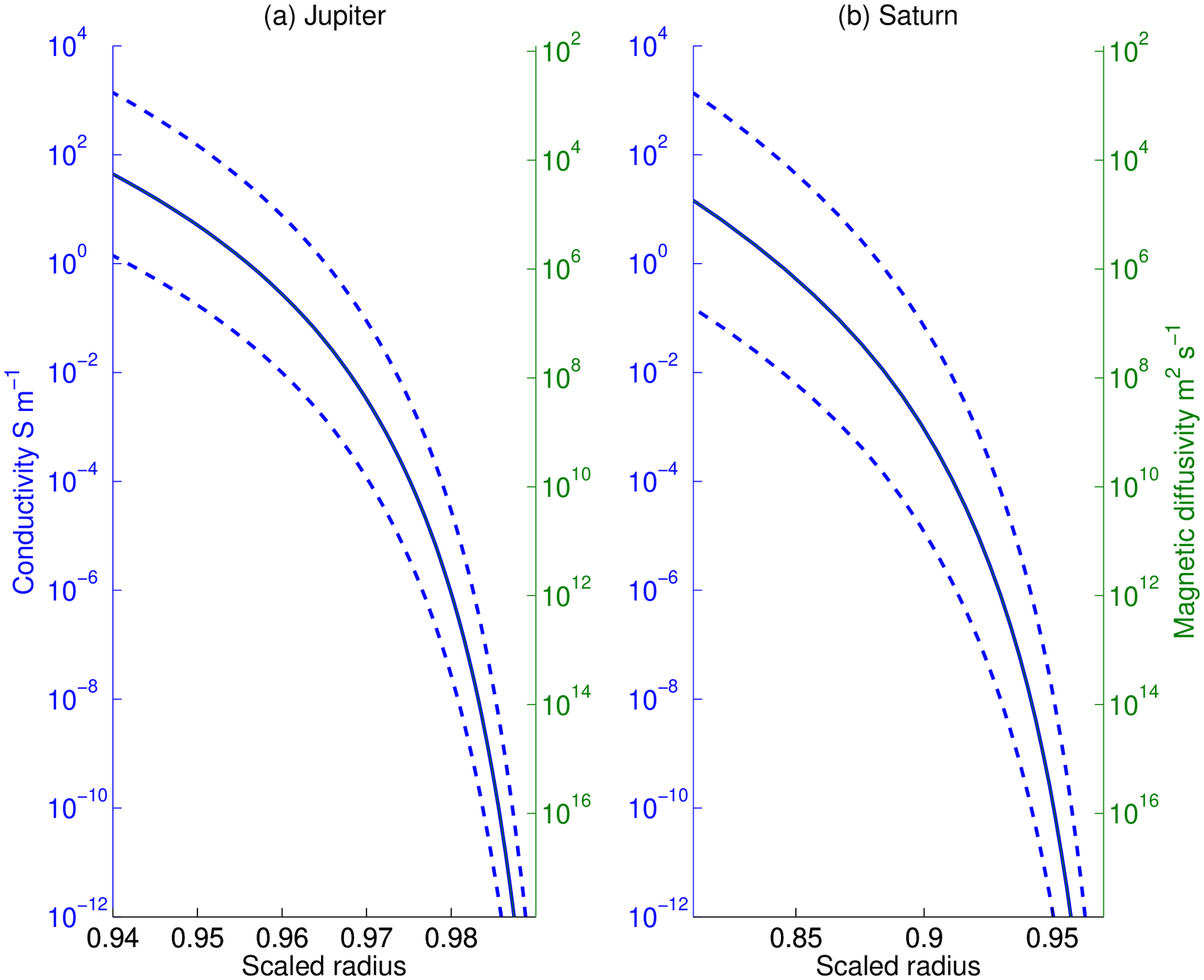} }
\caption{Electrical conductivity and magnetic diffusivity
distributions inside giant planets: (a) Jupiter; (b) Saturn. Values
of conductivity and magnetic diffusivity are plotted in the left and
right panels, respectively. Solid lines depict mean value; dashed
lines bound the range of uncertainties. } \label{config:
conductivity}
\end{figure}

\begin{figure}
\centerline{\includegraphics[width=5.2in]{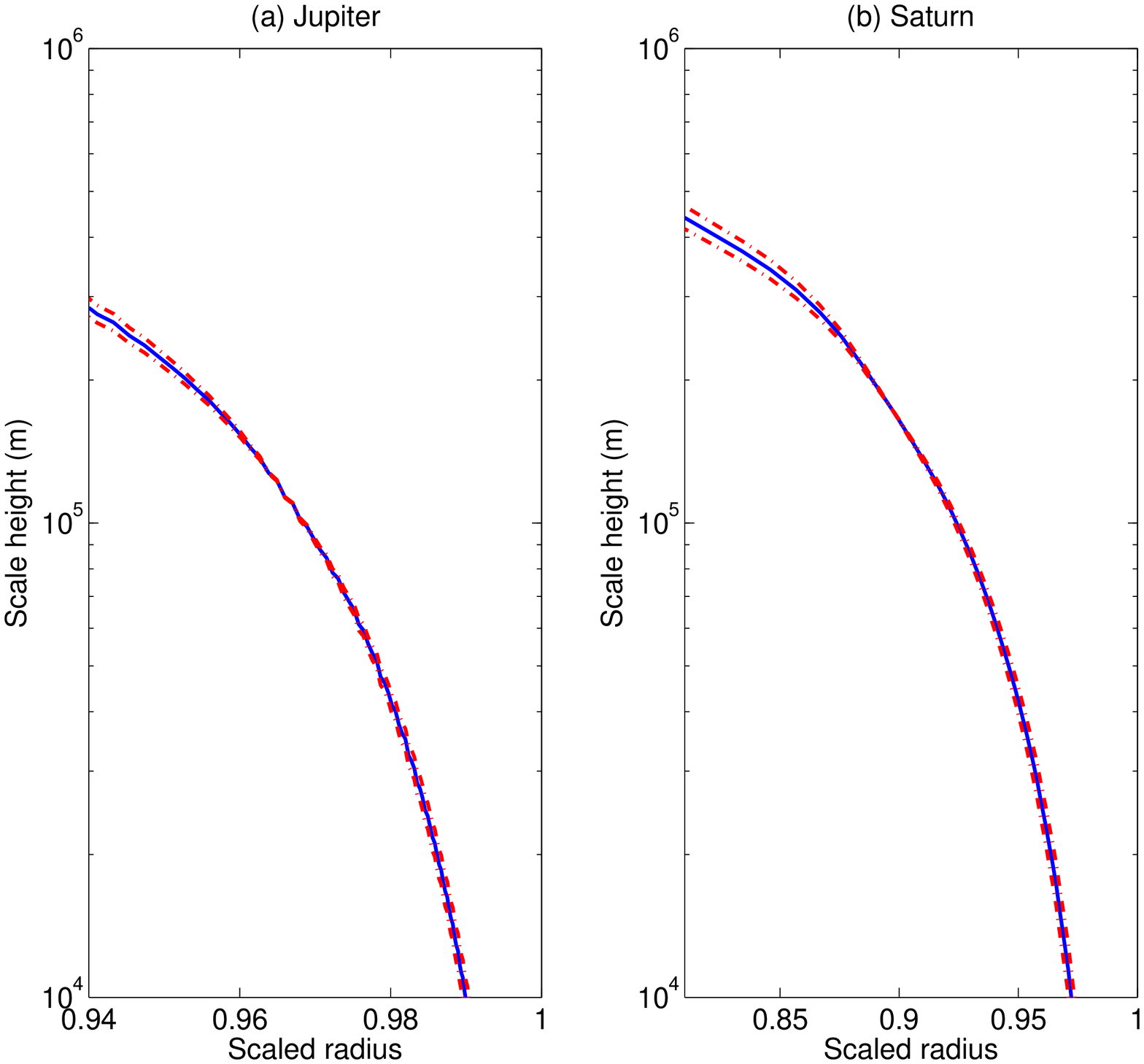} }
\caption{ Scale height of magnetic diffusivity as a function of
scaled radius: (a) Jupiter; (b) Saturn.} \label{config: scaleheight}
\end{figure}

In magnetohydrodynamics it is conventional to characterize the
electrical conductivity $\sigma$ in terms of the magnetic
diffusivity $\lambda = (\mu_0 \sigma)^{-1}$, where $\mu_0$ is the
magnetic permeability. Figure (\ref{config: conductivity}) shows
that the electrical conductivity of hydrogen decreases exponentially
outward from the metallic conducting region.  Therefore, the
magnetic diffusivity increases exponentially outward. The scale
height of magnetic diffusivity
\begin{equation}
H_{\lambda}(r) = \frac{\lambda(r)}{d \lambda(r) /dr}\,
\end{equation}
is shown in figure (\ref{config: scaleheight}).

\section{Ohmic Dissipation Based On Inward Extrapolation Of The
External Magnetic Field}
\label{sec:extrapolation}

We approximate the planet's magnetic field as axisymmetric;
Jupiter's dipole tilt is about $10^{\circ}$ and Saturn's less than
$0.1^{\circ}$ \citep{connerney93}. Then we evaluate the azimuthal
component of the magnetic field produced by differential rotation
acting on the poloidal components. The maximum penetration depth is
that of the level above which the associated Ohmic dissipation
matches the planet's net luminosity, ${\cal L}$.

To proceed, we need to know the poloidal magnetic field above the
maximum penetration depth. Here we assume that it can be determined
by inward extrapolation of the planet's external magnetic field.
This assumption is appropriate provided the magnetic Reynolds number
based on the convective velocity field, $R_m^c$, remains small down
to the maximum penetration depth, which our estimates suggest it
does.\footnote{An axisymmetric poloidal field is invariant under
differential rotation.}

Lack of accurate magnetic field measurements at high latitudes close
to Jupiter and Saturn makes the inward extrapolation of their
external magnetic fields somewhat uncertain. Thus we cannot exclude
the possibility that where $R_m^c\gg 1$ the magnetic field might be
closely aligned with the rotation axis. This possibility is examined
in \S \ref{sec:cylindrical}.

\subsection{Derivation of Ohmic dissipation}
\label{sec:detail:part1}

The time evolution of the magnetic field satisfies
\begin{equation}
\frac{\de \bb}{\de t} = \bn \times \left(\bu \times \bb \right) - \bn
\times \left[ \lambda(r) \bn \times \bb \right] \, ,
\label{eq:induc}
\end{equation}
where $\bu$ and $\bb$ denote velocity and magnetic field. We work in
spherical coordinates and set $\bu=U_\phi\ep=r\sin\theta\,\Omega
\ep$. The generation of toroidal field from poloidal field is
described by
\begin{align}
\frac{\partial B_\phi}{\partial
t}=r\sin\theta\left(\frac{\partial\Omega}{\partial
r}B_r+\frac{1}{r}\frac{\partial\Omega}{\partial\theta}B_\theta\right)
\nonumber
\\
+\frac{1}{r}\frac{\partial}{\partial
r}\left(\lambda\frac{\partial}{\partial
r}\left(rB_\phi\right)\right) \label{eq:phicomponent}
\\
+\frac{\lambda}{r^2}\frac{\partial}{\partial\theta}
\left(\frac{1}{\sin\theta}\frac{\partial}
{\partial\theta}\left(\sin\theta B_\phi\right)\right)\, . \nonumber
\end{align}

We seek a steady state solution noting that $B_\phi$ scales
proportional to $\lambda^{-1}$, and $H_\lambda$ is much smaller than
the length scale for the meridional variation of $U_\phi$ and $B$.
Thus we neglect $r^{-1}\partial/\partial\theta$ with respect to
$\partial/\partial r$, which is equivalent to assuming that $\bj_r
\ll \bj_\theta$.\footnote{A toy problem illustrating the effects
that $H_\lambda\ll H_\rho$ has on $\bj$ is presented in the
Appendix.} Then, by integrating the steady state version of equation
(\ref{eq:phicomponent}), we arrive at
\begin{equation}
\left[\lambda(r)\frac{\partial}{\partial
r}\left(rB_\phi\right)\right]_r^R\approx -\sin\theta\int_r^R
dr'r'^2\left(\frac{\partial\Omega}{\partial
r'}B_r+\frac{1}{r'}\frac{\partial\Omega}{\partial\theta}B_\theta\right)\,
. \label{eq:basic}
\end{equation}

With axial symmetry,
\begin{equation}
j_\theta=-\frac{1}{\mu_0 r}\frac{\partial}{\partial
r}\left(rB_\phi\right)\,  ,
\end{equation}
so
\begin{equation}
j_\theta(r, \theta)=\frac{-\sin\theta}{\mu_0 r \lambda(r)}\int_r^R
dr'r'^2\left(\frac{\partial\Omega}{\partial
r'}B_r+\frac{1}{r'}\frac{\partial\Omega}{\partial\theta}B_\theta\right)
+\frac{R\lambda(R)}{r\lambda(r)}j_\theta(R,\theta)\, .
\label{eq:current}
\end{equation}
Thus $j_\theta(r,\theta)$ is determined up to an unknown function of
$\theta$, namely $j_\theta(R,\theta)$, the current density in the
ionosphere.\footnote{$\lambda(r)$ is extremely large, effectively
infinite, in the neutral atmosphere but decreases dramatically in
the ionosphere.} There is a simpler and more intuitive way to derive
equation (\ref{eq:current}).  Start from $\bn\times \be=0$ and
$\bff=\be+\bu\times\bb$. Express the former in terms of a line
integral around a circuit consisting of two small arcs of the same
angular width $\delta\theta$, one at $r$ and the other at $R$,
connected by radial segments of length $\delta r=R-r$. Eliminate the
components of $\be$ in terms of those of $\bff$ and $\bu\times\bb$.
Then use $\bj=\sigma \bff$ to replace the components of $\bff$ in
terms of those of $\bj$. Make the assumption $|j_r|\ll|j_\theta|$
and equation (\ref{eq:current}) is derived.
\\

Next, we estimate the size of the rhs of equation
(\ref{eq:current}). Let $T_1$ and $T_2$ denote the first and second
terms, respectively. Steady state, axisymmetric currents cannot
close within the ionosphere. Nor can they penetrate inside the
planet due to the extremely low conductivity of its neutral
atmosphere. However, they can and do flow along field lines into the
magnetosphere. These currents produce torques which transfer angular
momentum from the planet's spin to plasma that is drifting outward
in the magnetosphere. In so doing, they tend to maintain that plasma
in approximate corotation with the planet. The torque, and hence
$j_\theta(R, \theta)$, are proportional to $\Omega{\dot M}$, where
$\dot M$ is the rate at which plasma is expelled from the
magnetosphere. Because the zonal winds cause $\Omega$ to vary by
only a few percent, $j_\theta(R,\theta)$ is a weak function of
$\theta$.  Thus, to a first approximation, $T_1$ and $T_2$ are
uncorrelated, so the magnitude of $j_\theta(r,\theta)$ is that of
the larger of these terms. We take a conservative approach and
accept the well-determined value of the first term as the minimum
value for $|j_\theta(r,\theta)|$. This is equivalent to treating the
ionosphere as an equipotential surface in the reference
frame rotating with planet's mean angular velocity.  \\

Ohmic dissipation per unit volume is equal to the square of the
current density divided by the electric conductivity.  Since
$j_\theta$ is dominant, we apply equation (\ref{eq:current}) to
obtain the total Ohmic dissipation above radius $r$;
\begin{equation}
P \approx  \frac{2 \pi}{\mu_0} \int_r^R \frac{dr'}{\lambda (r')}
\int_0^{\pi} d \theta \sin^3 \theta \left[\int_{r'}^R
dr''r''^2\left(\frac{\partial\Omega}{\partial
r''}B_r+\frac{1}{r''}\frac{\partial\Omega}{\partial\theta}B_\theta\right)\right]^2
\, . \label{eq:Ptot}
\end{equation}
For $\Omega$ constant on cylinders, the term in round brackets
reduces to $(\partial\Omega/\partial\varpi'')B_\varpi$, where
$\varpi=r\sin\theta$. Thus Ohmic dissipation vanishes both for solid
body rotation and for a poloidal field aligned parallel to the
rotation axis.

\subsection{Estimation of Ohmic dissipation inside Jupiter and Saturn.}
\label{odsubsec:ohmic:part1}

We apply equation (\ref{eq:Ptot}) to evaluate the total Ohmic
dissipation above radius $r$. Atmospheric zonal flows observed on
Jupiter and Saturn \citep{porco03,porco05} are taken to be constant
on cylinders outside a spherical truncation radius and to vanish
inside. Since the observed zonal flows are not exactly N-S
symmetric, we construct N-S symmetric profiles by reflecting the
northern hemisphere zonal flow about the equator.\footnote{We have
verified that using the reflected southern hemisphere zonal flow
makes a negligible difference to our results.} The magnetic fields
of Jupiter and Saturn have been measured by various spacecrafts and
fit by models dominated by a dipole plus smaller quadrupole and
octupole components \citep{connerney93}. We
adopt the axisymmetric part of these field models in our calculations.\\

\begin{figure}
\centerline{\includegraphics[width=5.2in]{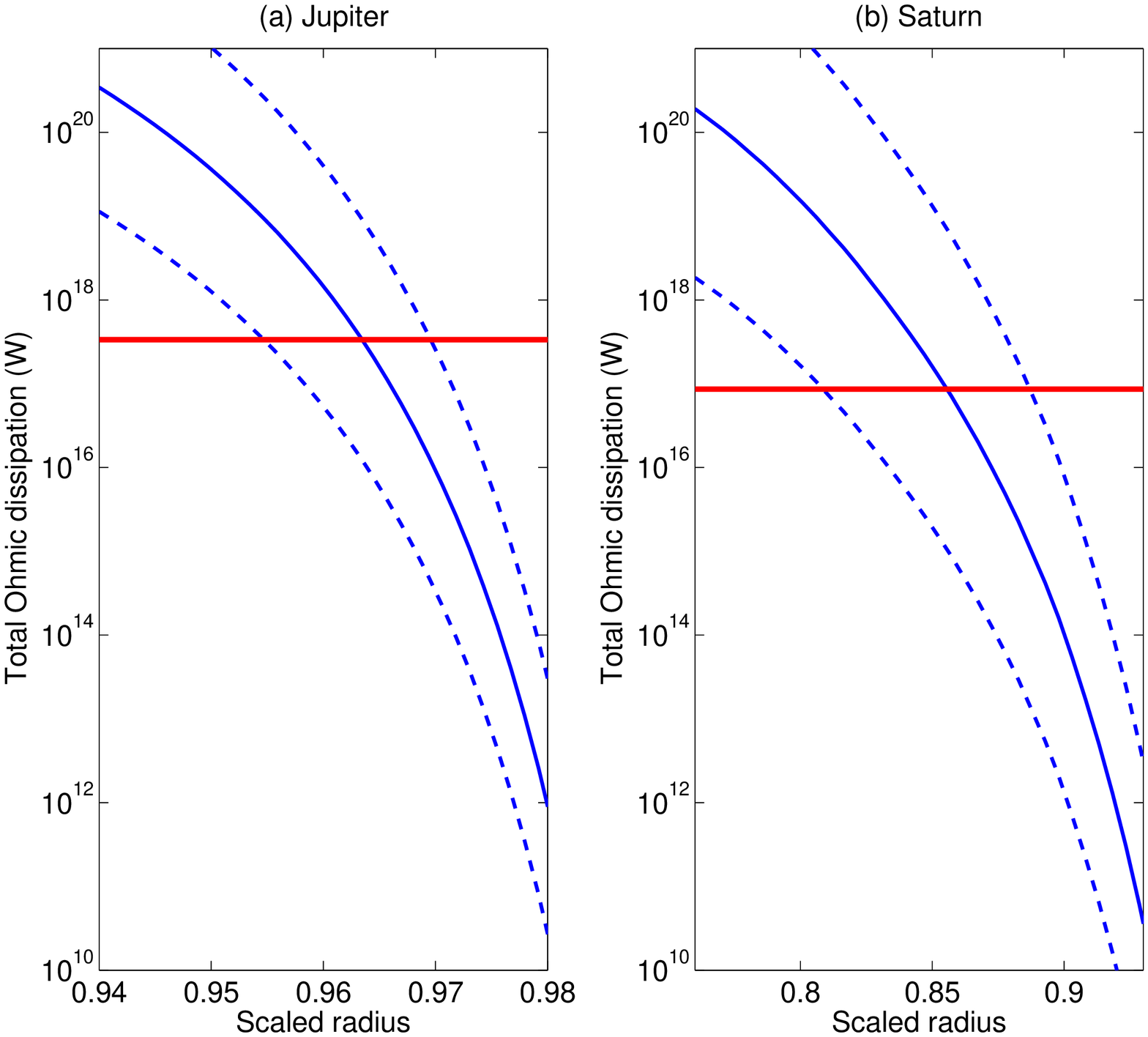} } \caption{
Observed zonal flow is taken to penetrate along cylinders until it
is truncated at $r_{mp}$. Solid curves display nominal values for
the total Ohmic dissipation, $P$, as a function of $r_{mp}/R$.
Dashed curves bound its uncertainty. Horizontal solid line marks the
planet's net luminosity, ${\cal L}$. At $r_{mp}$, $P={\cal L}$. }
\label{odfig: js_dissipation}
\end{figure}

Total Ohmic dissipation is plotted against truncation radius in
Figure (\ref{odfig: js_dissipation}) for Jupiter and Saturn. It
matches the planet's net luminosity at radii of $0.96 \ R_J$ and
$0.86 \ R_S$. The magnetic diffusivity at the radius of maximum
penetration is $10^7 \m^2 \s^{-1}$ for Jupiter and $3 \times 10^6
\m^2 \s^{-1}$ for Saturn. By comparison, the magnetic diffusivity is
at about $4\m^2 \s^{-1}$ at the planet's outer metallic cores
located at $0.84 R_J$ and $0.63 R_S$, respectively.

The magnitudes of the induced toroidal magnetic field and the
associated poloidal current are each inversely proportional to
$\lambda$ and thus increase inward. In figure (\ref{odfig:
js_toroidal_field}) we display the toroidal magnetic field as a
function of co-latitude at the maximum penetration depth. It reaches
a magnitude of about 15.0 G for Jupiter and about 0.5 G for Saturn.

The above estimates are based on the downward continuation of the
observed poloidal magnetic field. This is a reasonable procedure
provided the magnetic Reynolds number based on the convective
velocity is much smaller than unity above the radius of maximum
penetration.

\begin{figure}
\centerline{\includegraphics[width=5.2in]{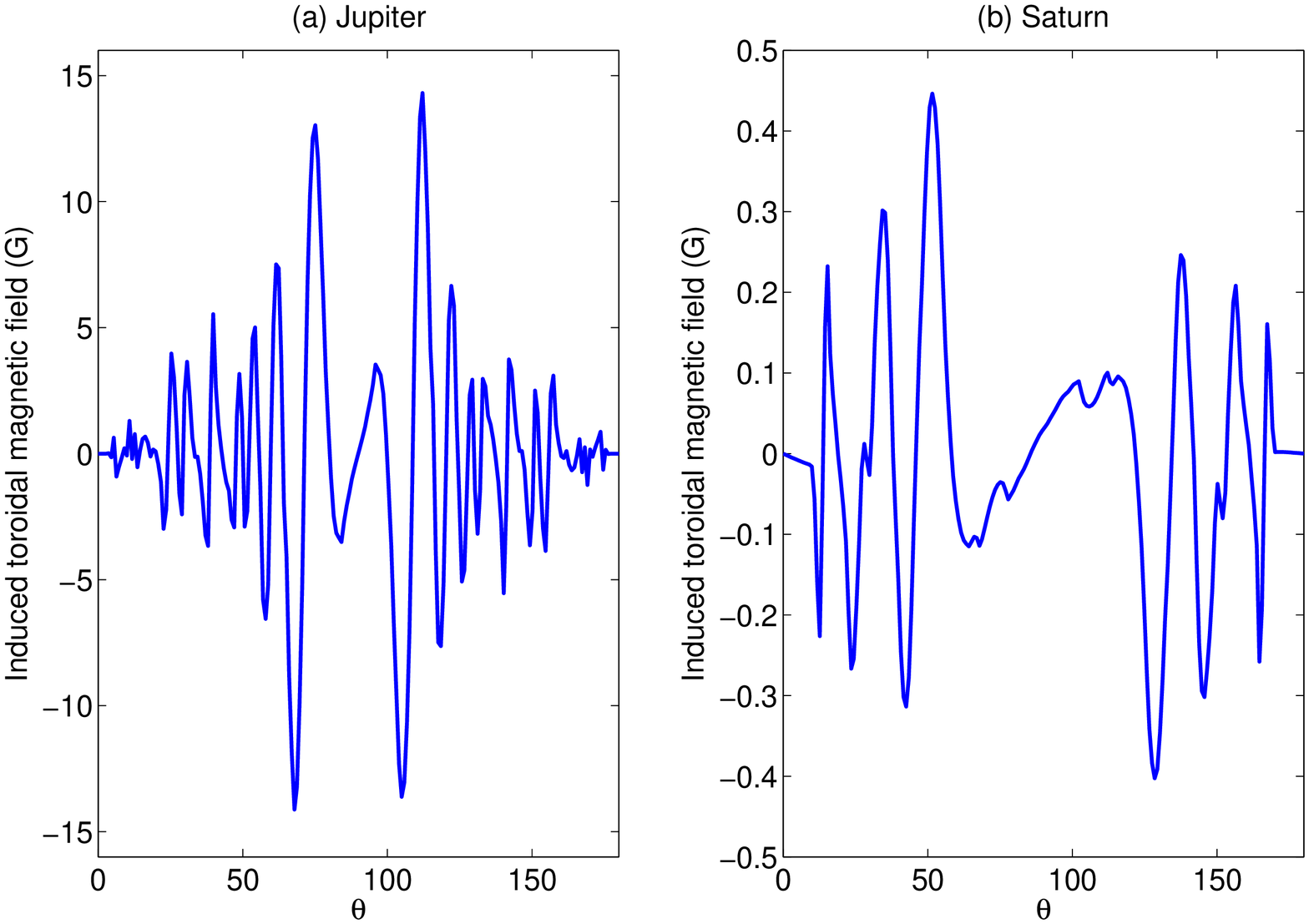} }
\caption{Toroidal magnetic field at the maximum penetration depth as
a function of colatitude: (a) Jupiter, (b) Saturn. } \label{odfig:
js_toroidal_field}
\end{figure}

\section{Ohmic Dissipation For Poloidal Field Lines
Aligned With The Rotation Axis} \label{sec:cylindrical}

Magnetic fields produced in some dynamo simulations show significant
alignment of poloidal field lines parallel to the rotation axis
\citep{glatzmaier05}. Such alignment would reduce Ohmic dissipation
associated with differential rotation since this dissipation is
proportional to $(d\Omega/d\varpi)B_\varpi$. However, alignment can
only occur in regions where $R_m^c\gtrsim 1$, so its overall effect
on the maximum penetration depth of atmospheric zonal winds on
Jupiter and Saturn is not obvious.

\begin{figure}
\centerline{\includegraphics[width=8.0in]{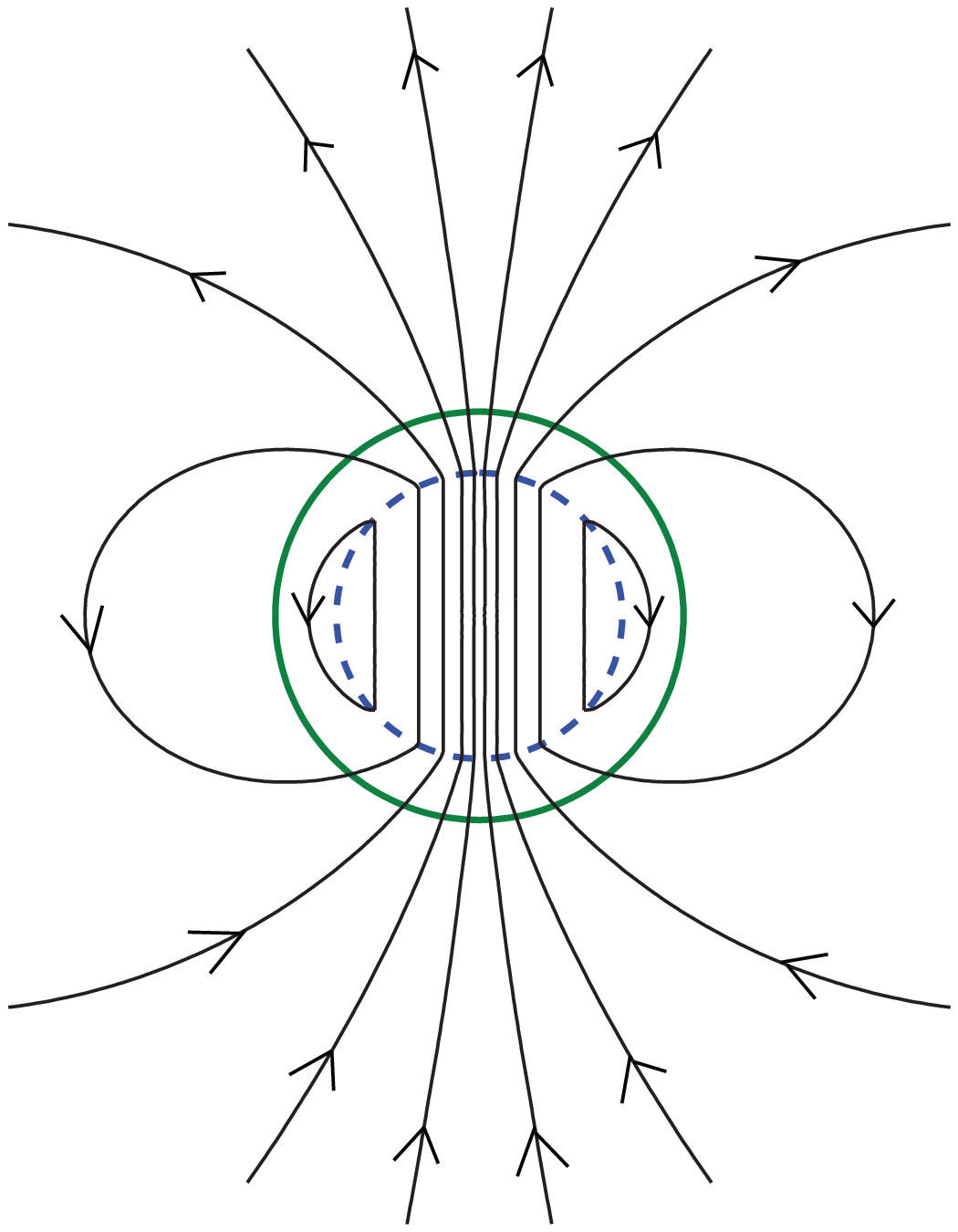} }
\caption{Poloidal magnetic field lines for $p=5$.  }
\label{fig_mag}
\end{figure}

To examine the effects of alignment on Ohmic dissipation, we
consider a model in which the magnetic field is perfectly aligned
inside a sphere of radius $r_{*}<R$ (See figure \ref{fig_mag}),
\begin{equation}
B_\varpi=0 \quad {\rm and} \quad B_z =
B_0\left[1-\left(\varpi/r_*\right)^2\right]^{(p-1)}\, ,
\end{equation}
with $p$ a positive integer. The constant $B_0$ is set to match the
planet's external magnetic dipole. For larger $p$, the internal
magnetic field is more concentrated towards the rotation axis.

The spherical shell $r=r_*$ marks the outer boundary of the dynamo
region. For $r< r_*$,  $R_m^c> 1$ and the poloidal components of the
fluid motions and magnetic field are strongly coupled. We assume
that the dynamo maintains the aligned poloidal magnetic field
against Ohmic decay. For $r>r_*$, the poloidal magnetic field is
taken to be a potential field. Differential rotation acting on the
poloidal magnetic field produces a toroidal magnetic field as
described in \S \ref{sec:extrapolation}.

Outside $r_{*}$, $\bn\times\bb=0$,
\begin{eqnarray}
B_r &=& \sum_{n=1}^{\infty} 2n\left(R/r
\right)^{2n+1}P_{2n-1}^0(\cos\theta)g_{2n-1}
 \cr
B_\theta &=& -\sum_{n=1}^{\infty}\left(R/r \right)^{2n+1}
P_{2n-1}^1(\cos\theta)g_{2n-1} \, . \label{eq:Bext}
\end{eqnarray}

In order to match the internal field with the external field,
we expand the internal field into spherical
harmonics at $r_*$,
\begin{eqnarray}
B_r&=& B_0\sum_{n=1}^\infty
v_{2n-1}P_{2n-1}^0(\cos\theta) \cr
B_\theta&=&B_0\sum_{n=1}^\infty
w_{2n-1}P^1_{2n-1}(\cos\theta) \, , \label{eq:Bint}
\end{eqnarray}
with
\begin{eqnarray}
v_{2n-1}&=&\frac{(4n-1)}{2}\int_{-1}^{1} dx\,
x^{2p-1}P^0_{2n-1}(x)\cr
w_{2n-1}&=&-\frac{(4n-1)}{4(2n-1)n}\int_{-1}^{1} dx\,
x^{2p-1}(1-x^2)^{1/2}P^1_{2n-1}(x) \, . \label{eq:Bintcoef}
\end{eqnarray}
The radial component of the magnetic field is continuous across
$r_{*}$. Thus
\begin{equation}
g_{2n-1} = \left(r_*/R\right)^{2n+1}v_{2n-1}B_0/(2n)\, ,
\label{eq:gnvn}
\end{equation}
where $B_0=2(R/r_*)^3g_1/v_1$ in order that $g_1$ match the planet's
external dipole moment.

Ohmic dissipation comes from three sources: the surface current at
$r_*$, the current associated with the nonuniform field inside $r_*$
(for $p>1$), and the current which arises from interaction of the
vacuum field with the zonal flow outside $r_*$. We treat each of
these in turn.

\subsection{Dissipation due to surface current}

\begin{figure}
\centerline{\includegraphics[width=6.2in]{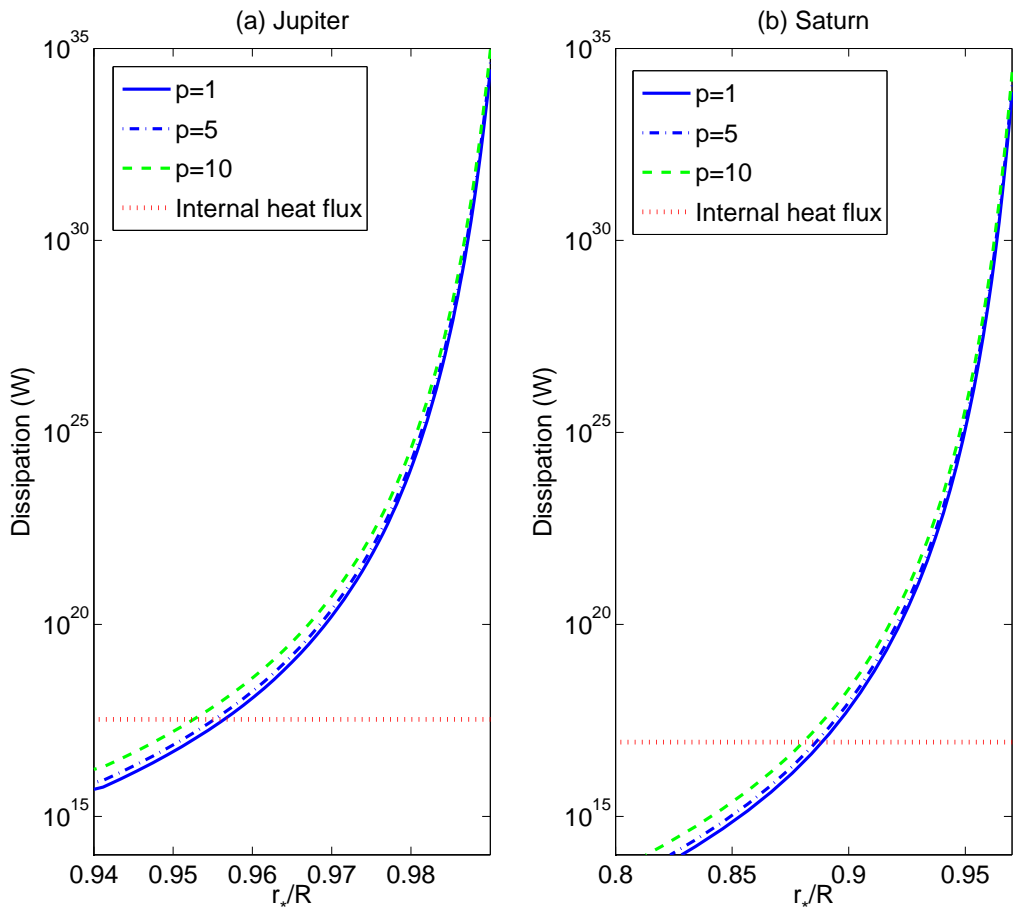} } \caption{
Ohmic dissipation due to surface current as a function of $r_*/R$.
(a) Jupiter; (b) Saturn.} \label{fig_sd}
\end{figure}

Because $B_\theta$ is discontinuous across $r_*$, the associated
surface current, $J_s = \Delta B_\theta/\mu_0$, would give rise to
infinite Ohmic dissipation. However, the transition between internal
and external field should be spread across a length scale of order
the scale height of the magnetic diffusivity, $H_\lambda$. Then
Ohmic dissipation from the surface current evaluates to
\begin{equation}
P \approx \frac{2 \pi r_*^2}{H_\lambda}
\frac{\lambda}{\mu_0}\int_{-1}^1 dx
 \left(\Delta B_\theta(x)\right)^2
= \frac{8n(2n-1)\pi r_*^2\lambda B_0^2}{(4n-1)\mu_0H_\lambda}
\sum_{n=1}^\infty\left(\frac{v_{2n-1}}{2n} + w_{2n-1}\right)^2 \, .
\end{equation}
Plots of Ohmic dissipation due to the surface current as a function
of $r_*$ for different values of $p$ are displayed in figure
(\ref{fig_sd}).

\subsection{Dissipation due to current inside $r_*$.}

\begin{figure}
\centerline{\includegraphics[width=5.2in]{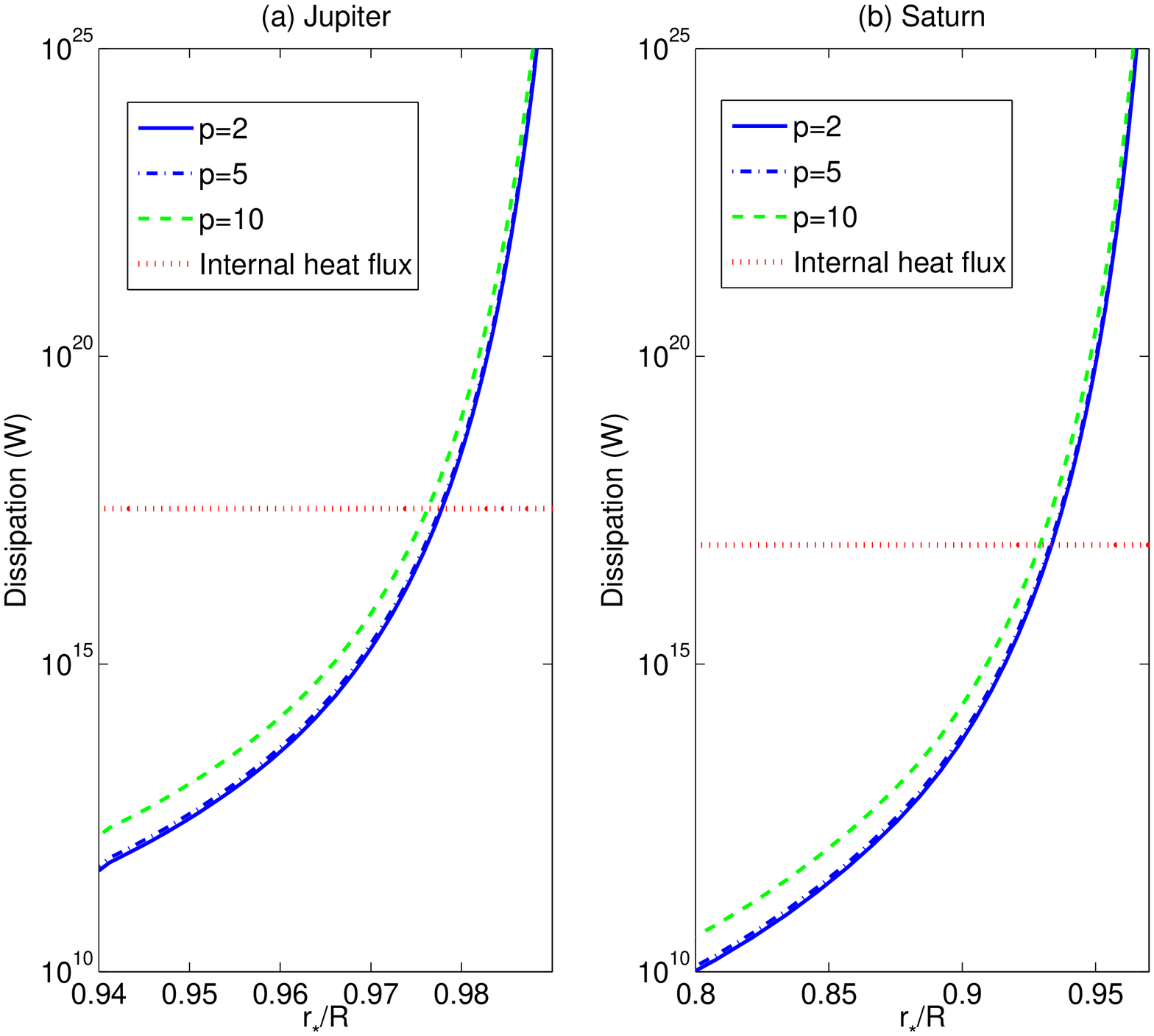} }
\caption{Ohmic dissipation due to internal current as a function of
$r_*/R$. (a) Jupiter; (b) Saturn.} \label{fig_sd_inter}
\end{figure}

The azimuthal current density
\begin{equation}
\jj_{\phi} = - \frac{1}{\mu_0} \frac{\de B_z}{\de
\varpi}= \frac{2(p-1)B_0}{\mu_0}\left(\frac{\varpi}{r_*^2}\right)\left(1-\frac{\varpi^2}{r_*^2}
\right)^{p-2} \,
\end{equation}
produces Ohmic dissipation in a layer of thickness $H_{\lambda}$
given by
\begin{equation}
P \approx  \frac{2\pi r_*^2 \lambda H_\lambda}{\mu_0} \int_0^\pi
\left(\frac{\de B_z}{\de \varpi} \right)^2 \sin\theta
d\theta=\frac{32\pi(p-1)^2\lambda H_\lambda
B_0^2}{(4p-5)(4p-7)\mu_0}\, . \label{eq:Pint}
\end{equation}
Figure (\ref{fig_sd_inter}) displays the internal dissipation as a
function of $r_*$ for different values $p$.

\subsection{Ohmic dissipation due to action of the zonal wind
on the poloidal magnetic field.}

\begin{figure}
\centerline{\includegraphics[width=6.2in]{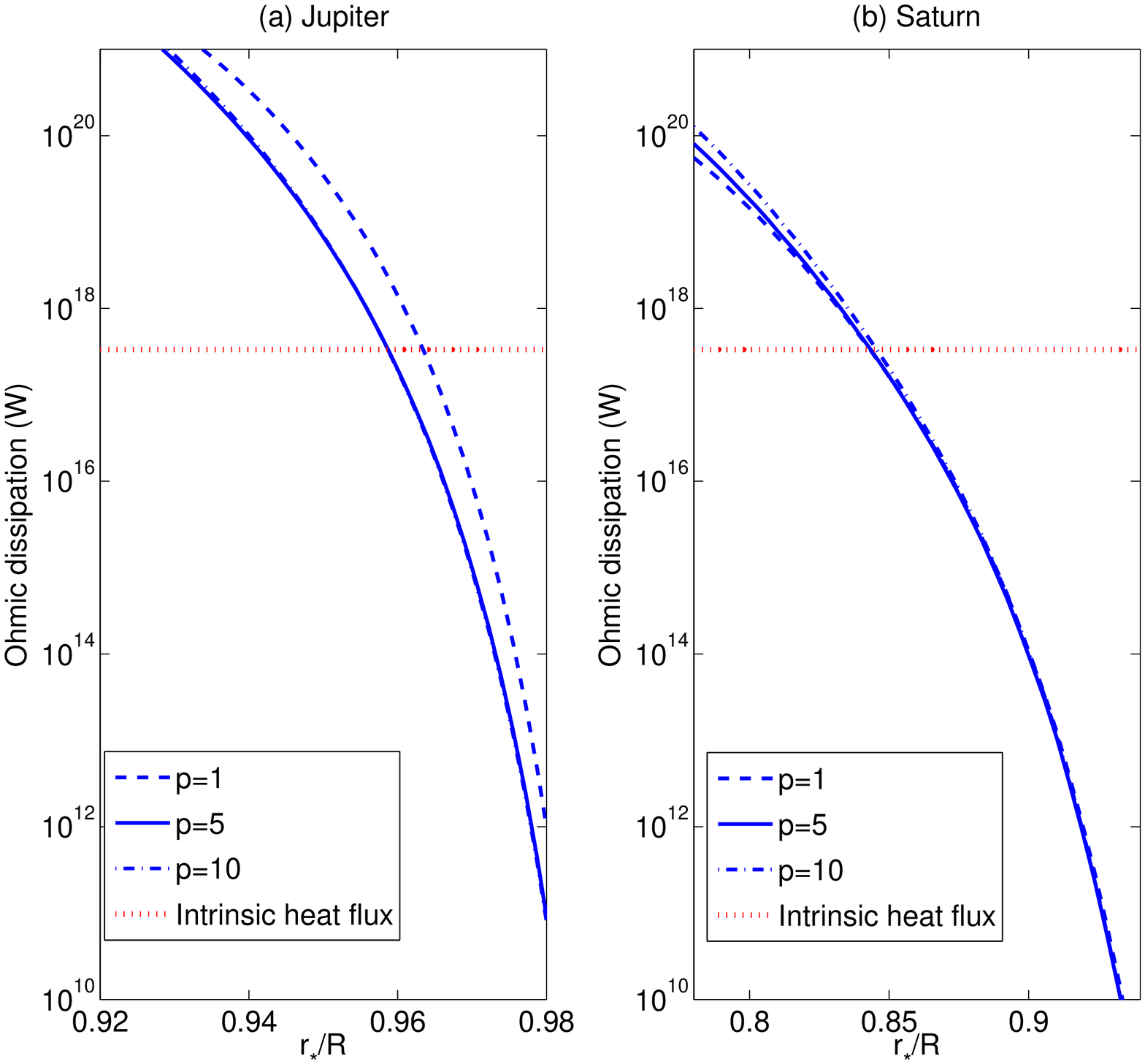} }
\caption{Ohmic dissipation due to interaction of $B_\varpi$ with
$\partial\Omega/\partial\varpi$ as a function of $r_*/R$. (a)
Jupiter; (b) Saturn.} \label{fig_oh_flow}
\end{figure}

$B_\varpi\neq 0$ for $r>r_*$, so Ohmic dissipation results from the
action of the zonal wind on the poloidal magnetic field. We evaluate
it with the aid of equation (\ref{eq:Ptot}) and plot the results in
figure (\ref{fig_oh_flow}).

\subsection{Total Ohmic dissipation}

\begin{figure}
\centerline{\includegraphics[width=6.2in]{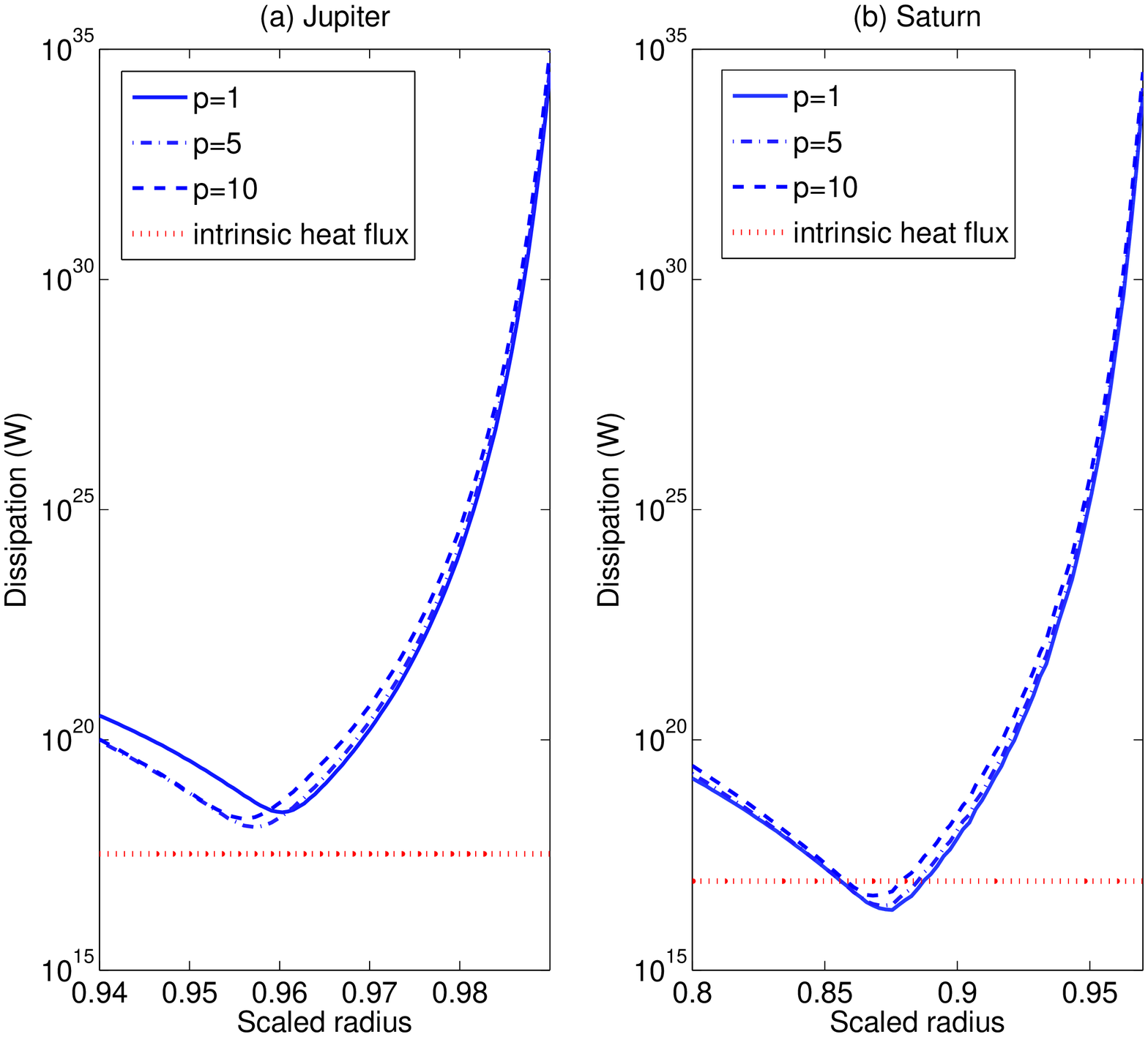} }
\caption{Total Ohmic dissipation as a function of $r_*/R$. (a)
Jupiter; (b) Saturn.} \label{fig:totaldiss}
\end{figure}

The total Ohmic dissipation from all three sources is plotted as a
function of $r_*$ in figure (\ref{fig:totaldiss}) for different
values of $p$. Comparison with figure (\ref{odfig: js_dissipation})
reveals that alignment does not significantly increase the maximum
penetration depth of zonal winds on Jupiter and Saturn beyond that
calculated in \S \ref{sec:extrapolation} by downward extrapolation
of their external magnetic fields.

The arguments in this section do not refer to the magnetic Reynolds
number, $R_m^c$, associated with the convective velocity. Thus the
maximum penetration depth we deduce is independent of this quantity.
Nevertheless, in reality, the model field we investigate only makes
physical sense if $R_m^c\gg 1$ at $r_*$.

\section{How Deep Do The Zonal Flows Penetrate?}
\label{sec:truncate}

Together, \S \ref{sec:extrapolation} and \S \ref{sec:cylindrical}
place an upper limit on the depth to which the zonal winds observed
in the atmosphere of a giant planet could penetrate. Here we address
the more difficult issue of how deeply they actually do penetrate.
In particular, do they extend unabated into the convective envelope.
It proves convenient to work in cylindrical coordinates $(\varpi,
\phi, z)$.

The Navier-Stokes equation which governs the motion of the fluid
reads
\begin{equation}
\frac{\de \bu}{\de t} + \left(\bu \cdot \bn\right)\bu + 2 \bOmega_P
\times \bu = -\frac{\bn p}{\rho} - \bn \Phi_{\rm tot} +
\frac{\left(\bn \times \bb \right) \times \bb}{\mu_0\rho}  \, ,
\label{eq:Navier}
\end{equation}
where $\bu$ is the total velocity measured relative to a frame
rotating at angular speed $\Omega_P$ (the assumed uniform angular
velocity of the planet's metallic core), $\rho$ is the density, $p$
is the pressure, and $\Phi_{\rm tot}$ is the gravitational plus
centrifugal potential.\footnote{We ignore the utterly negligible
viscous stress.} We express the steady-state limit of this equation
as
\begin{equation}
2 \bOmega_P \times \obu = -\frac{\bn \op}{\orho} - \bn
{\overline\Phi}_{\rm tot} + \oba\, , \label{eq:NavierSS}
\end{equation}
where the overbar denotes time average and
\begin{equation}
\oba= \frac{\overline{\left(\bn \times \bb \right) \times
\bb}}{\mu_0\orho} -{\overline{\left(\bu \cdot \bn\right)\bu}} \, ,
\label{eq:defa}
\end{equation}
Smaller terms that involve fluctuations of $\rho$, $p$, and
$\Phi_{\rm tot}$ have been discarded. Individually, the first and
second terms on the rhs of equation (\ref{eq:NavierSS}) are larger,
by far, than the others. Thus
\begin{equation}
\frac{\bn \op}{\orho} \approx -\bn {\overline\Phi}_{\rm tot}\, .
\label{eq:equilbrium}
\end{equation}

Taking the curl of equation (\ref{eq:NavierSS}), we arrive at
\begin{equation}
2\Omega_P\frac{\partial \out}{\partial z}=-\left.
\frac{\partial\ln\rho}{\partial s}\right|_p \left(\bn\os\times
\obg\right)\cdot\ep-\left(\bn\times \oba\right)\cdot\ep    \, ,
\label{eq:curlss}
\end{equation}
with $\obg=-\bn\oPhi_{\rm tot}$. Thus $\partial\out/\partial z=0$
for an uniform composition isentrope in the absence of Reynolds and
Maxwell stresses. This is the Taylor-Proudman state. Next we bound
departures from this state that each term on the rhs of equation
(\ref{eq:curlss}) could produce.

\subsection{Buoyancy}

\subsubsection{In the convective envelope}

Here we are entering uncertain territory. Convection transports the
net luminosity in the fluid interior, but our understanding of
turbulent convection is limited even for nonrotating systems.
Rotation and especially strong differential rotation add additional
complexity. It is not obvious which of the terms on the rhs of
equation (\ref{eq:curlss}) is dominant for conditions appropriate to
a convective envelope.

We are guided by \citet{ingersoll82} who model convection under
conditions of strong differential rotation which shears convective
cell in the azimuthal direction. They argue that under these
conditions, the magnitude of the component of $\bn\os$ along $\obg$
must satisfy
\begin{equation}
g\left|\frac{\partial\ln\rho}{\partial
 s}\right|_p\frac{|\bn \os\cdot\obg|}{|\obg|}
\sim \left(\frac{\de U_{\phi}}{\de \varpi}\right)^2\, ;
\label{eq:case2gradS}
\end{equation}
In other words, the Richardson number based on the rate of shear is
of order unity.

The variation of $\out$ along $z$ depends upon the component of
$\bn\os$ that is orthogonal to both $\obg$ and $\ep$ about which
mixing length models are silent. We parameterize this component in
terms of $\bn \os\cdot{\hat{\obg}}$ and the angle $\delta$ between
$\bn\os$ and $\obg$. Buoyancy drives convection, so it is to be
expected that $\delta\ll 1$. Combining equations (\ref{eq:curlss})
and (\ref{eq:case2gradS}) yields
\begin{equation}
\left|\frac{H_\rho}{U_{\phi}}\frac{\de U_{\phi}}{\de z}\right|\sim
\frac{H_\rho}{2\Omega_P U_{\phi}}\left(\frac{\de U_{\phi}}{\de
\varpi}\right)^2\tan\delta  \, . \label{eq:case2}
\end{equation}
Since $\obg$ is approximately aligned along the {\it spherical}
radial direction, $|\bn \os|\tan\delta$ is essentially the magnitude
of latitudinal component of $\bn \os$.

At the maximum penetration depth,
\begin{equation}
\left|\frac{H_\rho}{\out}\frac{\de \out}{\de z}\right| \sim
0.6\tan\delta
\end{equation}
for Jupiter, and
\begin{equation}
\left|\frac{H_\rho}{\out}\frac{\de \out}{\de z}\right| \sim
0.3\tan\delta\, .
\end{equation}
for Saturn. As stated above, we expect that $\tan\delta\ll 1$.
Moreover, its numerical coefficients is proportional to $H_\rho$,
which decreases outward. Thus buoyancy is unlikely to effect a
significant departure from the Taylor-Proudman state in the
convective envelope.

\subsubsection{In the radiative atmosphere}

There are conflicting views about the strength of the static
stability in the radiative atmospheres of Jupiter and Saturn and the
depth to which stable layers extend. Here we consider how
Taylor-Proudman columns might be truncated in strongly stable layers
in which $s$ changes on the same {\it spherical} radial scale as
$\rho$. In such layers
\begin{equation}
\left|\frac{\partial\ln\rho}{\partial s}\right|_p H_\rho|\bn s| \sim
1\, \label{eq:fracdelUrad}
\end{equation}
in equation (\ref{eq:curlss}) to obtain
\begin{equation}
\left|\frac{H_\rho}{\out}\frac{\de \out}{\de z}\right|
\sim\frac{g}{2\Omega_P \out}\tan{\delta}\, .\label{eq:finalatmos}
\end{equation}

Numerical evaluation yields
\begin{equation}
\left|\frac{H_\rho}{\out}\frac{\de
 U_{\phi}}{\de z}\right|
\sim 600 \tan{\delta} \, . \label{eq:finalatmos_Jupiter}
\end{equation}
for Jupiter, and
\begin{equation}
\left|\frac{H_\rho}{\out}\frac{\de \out}{\de z}\right| \sim 200
\tan{\delta} \, . \label{eq:finalatmos_Saturn}
\end{equation}
for Saturn. Taylor-Proudman columns could be truncated in the
radiative atmosphere provided surfaces of constant entropy were
inclined to those of constant potential in the latitudinal direction
by more than a fraction of a degree. The resulting velocity field
would constitute a strong thermal wind.

\subsection{Magnetic stresses}

Deviations from the Taylor-Proudman state caused by Maxwell stresses
follow from
\begin{equation}
\left|\frac{H_\rho}{\out}\frac{\de \out}{\de z}\right|=
\frac{H_\rho}{2\mu_0\Omega_P \left|\out\right|}
\left|\ep\cdot\bn\times\left(\frac{\overline{\left( \bn \times \bb
\right) \times \bb}}{\orho}\right)\right| \sim
\frac{H_\rho\overline{ B^2}}{2\mu_0\Omega_P|\out|\orho\ell^2}\, ,
\label{eq:tpmag}
\end{equation}
where $\ell$ is the typical scale over which $\bb$ varies. The order
of magnitude estimate assumes $\ell\lesssim H_\rho$.

Next we bound $\overline{B^2}/\ell^2$ by applying the Ohmic
constraint. The latter reduces to
\begin{equation}
\frac{4\pi\lambda R^2 H_\rho\overline{B^2}}{\mu_0\ell^2}\lesssim
{\cal L}\, . \label{eq:ohmfluc}
\end{equation}
Eliminating $\overline{B^2}/\ell^2$ between equations
(\ref{eq:tpmag}) and (\ref{eq:ohmfluc}), we arrive at
\begin{equation}
\left|\frac{H_\rho}{\out}\frac{\de \out}{\de z}\right|\lesssim
\frac{{\cal L}}{8\pi\lambda\Omega_P|\out|\rho R^2} \, .
\label{eq:magbound}
\end{equation}
Numerical evaluation yields
\begin{equation}
\left|\frac{H_\rho}{\out}\frac{\de \out}{\de z}\right|\lesssim
10^{-5} \label{eq:nummaxwell}
\end{equation}
at the maximum penetration depths in Jupiter and Saturn. Moreover,
this ratio decreases sharply outward. Clearly, magnetic stresses are
incapable of truncating the observed zonal flows.

\subsection{Reynolds stresses}

Departures from the Taylor-Proudman state caused by Reynolds
stresses obey
\begin{equation}
\left|\frac{H_\rho}{\out}\frac{\de \out}{\de
z}\right|=\frac{H_\rho}{2\Omega_P|\out|}
\left|\ep\cdot\bn\times\left(\overline{\left(\bu
\cdot \bn\right)\bu}\right)\right|\, . \label{eq:tprey}
\end{equation}
We treat separately contributions from the mean axisymmetric flow,
$\obu$, and from fluctuations about it, $\blu=\bu-\obu$.

The sole contribution involving $\out$ can be absorbed by adding
$\out/\varpi$ to $\Omega_P$ on the lhs of equation
(\ref{eq:NavierSS}) and in what follows. We do not consider it
further. The poloidal part of $\obu$, denoted by $\obu_p$, describes
meridional circulation. It yields
\begin{equation}
\left|\frac{H_\rho}{\out}\frac{\de \out}{\de z}\right|\sim
\frac{\overline{U^2_p}}{2\Omega_P|\out|H_\rho}\, , \label{eq:reymer}
\end{equation}
where the variation scale of $\obu_p$ is set to $H_\rho$. With
parameters appropriate to Jupiter and Saturn, the rhs of equation
(\ref{eq:reymer}) is $\ll 1$ for $|\overline {U_p}|\ll |\out|$.

Velocity fluctuations contribute
\begin{equation}
\left|\frac{H_\rho}{\out}\frac{\de \out}{\de z}\right|\sim
\frac{H_\rho \overline{u_\phi^2}}{2\Omega_P|\out|\ell_p\ell_\phi}\,
, \label{eq:reyfluc}
\end{equation}
where $\ell_p$ and $\ell_\phi$ are the meridional and azimuthal
scales of the convective cells. In deriving equation
(\ref{eq:reyfluc}), we note that that azimuthal stretching of eddies
by strong differential rotation results in $\ell_\phi\gtrsim\ell_p$
and mass conservation implies $u_p\ell_\phi\sim u_\phi\ell_p$. To
bound $\overline{u_\phi^2}$, we follow \citet{ingersoll82} and adopt
$|\partial\out/\partial \varpi|(\ell_p/\ell_\phi)$ as the convective
rate. In each scale height, turbulent mechanical energy is
dissipated as heat at a rate $\sim {\cal L}$. Hence
\begin{equation}
4\pi R^2H_\rho\orho\overline{u_\phi^2}\frac{\partial\out}{\partial
\varpi}\frac{\ell_p}{\ell_\phi}\sim {\cal L}\, . \label{eq:carnot}
\end{equation}
Together, equations (\ref{eq:reyfluc}) and (\ref{eq:carnot}) yield
\begin{equation}
\left|\frac{H_\rho}{\out}\frac{\de \out}{\de z}\right|\sim
\frac{{\cal L}}{8\pi\Omega_P|\out||\partial\out/\partial
\varpi|\orho R^2\ell_p^2}\lesssim  \frac{{\cal
L}}{8\pi\Omega_P\out^2\orho R^3\Delta\theta}\, , \label{eq:reydev}
\end{equation}
where $\Delta\theta$ denotes the typical latitudinal width of the
zonal jets. In accord with \citet{ingersoll82}, we set
$|\partial\out/\partial \varpi|\sim |\out|/\ell_p$ in arriving at
the final form of equation (\ref{eq:reydev}). Numerical evaluation
with parameters appropriate to the tops of the convection zones in
Jupiter and Saturn gives
\begin{equation}
\left|\frac{H_\rho}{\out}\frac{\de \out}{\de z}\right| \lesssim
10^{-5}\,  .\label{eq:gradUReynum}
\end{equation}

\subsection{Maximum width of an equatorial jet}
\label{odsubsec: jetwidth} 
\begin{figure}

\centerline{\includegraphics[width=5.2in]{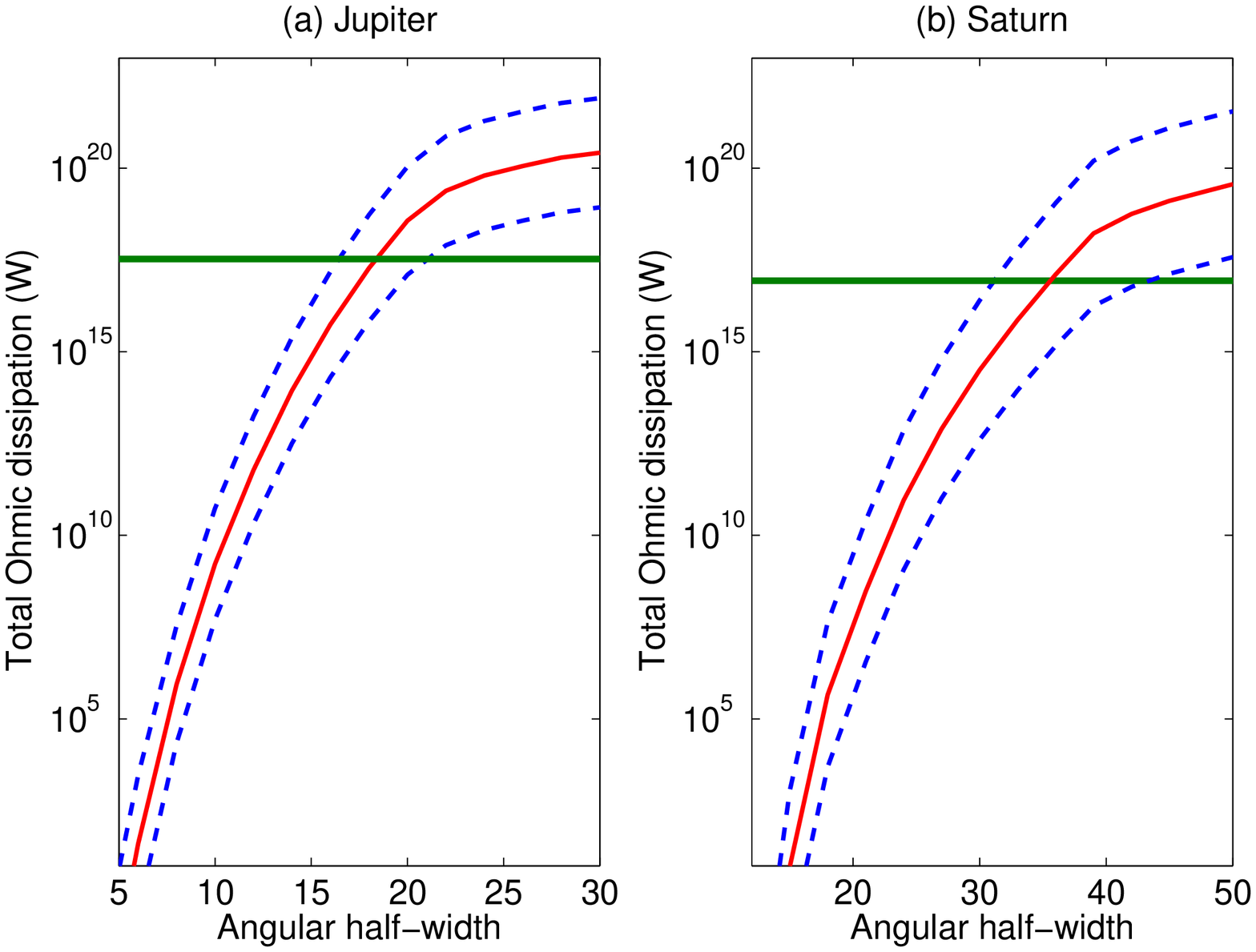} }
\caption{Total Ohmic dissipation rate verses jet's angular
half-width. Horizontal solid lines mark the planet's net luminosity.
Dashed lines indicate the range of uncertainties. Central value of
the maximum half-width is $18.5^{\circ}$ for Jupiter and
$35.5^{\circ}$ for Saturn.} \label{odfig: cut_off_angle}
\end{figure}

A sufficiently narrow equatorial jet
could maintain constant velocity on cylinders throughout the planet.
As an example, we consider the specific velocity profile
\begin{equation}
\out = U_{\phi 0} \sin \left(\frac{\pi}{2}
\frac{\left(\theta-\theta_0\right)}{\left(\pi/2-\theta_0\right)}
\right)^{\frac{1}{10}} \ \ \ \ \mathrm{if} \ \ \ \ \theta < \pi -
\theta_0;
\end{equation}
and,
\begin{equation}
\out = 0 \ \ \ \ \ \mathrm{if} \ \ \ \ \theta < \theta_0 \ \
\mathrm{and}
 \ \ \ \theta > \pi - \theta_0;
\end{equation}
so the jet has equatorial velocity $U_{\phi 0}$ and angular
half-width $\pi/2-\theta_0$. For Jupiter and Saturn, $U_{\phi 0}$ is
approximately $140 \m \s^{-1}$ and $400 \m \s^{-1}$, respectively.
Figure (\ref{odfig: cut_off_angle}) displays the calculated Ohmic
dissipation rate as a function of the jet half-width. The maximum
half-width is about $18.5^{\circ}$ for Jupiter, and $35.5^{\circ}$
for Saturn.  The maximum half widths are related to the fractional
radii of maximum penetration, $r_{mp}/R$, calculated in section
\ref{sec:extrapolation} by
\begin{equation}
\cos\theta_0\approx \frac{r_{mp}}{R}\, .
\end{equation}

\bigskip\bigskip\bigskip

\section{Discussion.}
\label{sec:discussion}

The condition that the total Ohmic dissipation not exceed the
planet's net luminosity sets an upper bound on the depth to which
the zonal flows observed in the atmospheres of Jupiter and Saturn
could penetrate. At these depths, the magnetic Reynolds number,
based on the observed zonal winds and the scale height of the
magnetic diffusivity, is of order unity.

We consider it unlikely that the observed flows extend to the depth
of maximum penetration because that would require the
Taylor-Proudman constraint to be violated in the convective
envelope. We have been unable to identify any plausible mechanism
that could do this. Lack of a rigorous model for convection is a
weakness. We base our analysis of the robustness of the
Taylor-Proudman constraint on the mixing length model designed by
\citet{ingersoll82} to apply under conditions of strong differential
rotation. However, we have confirmed that the conclusions we draw
from it also follow from other convection models that consider only
solid body rotation, or even those that neglect the effects of
rotation entirely.

The boundaries of the cylindrical extensions of the equatorial jets
essentially coincide with the maximum penetration depths. Thus these
jets could maintain constant velocities along cylinders through the
planets. Winds measured at $7.4^{\circ} N$ by the Galileo probe as
it descended into Jupiter are consistent with this possibility.

\paragraph{Acknowledgments} We are grateful for support from the NASA
Planetary Geology and Geophysics program.

\section{Appendix}
\label{sec:appendix}

\begin{figure}
\centerline{\includegraphics[width=3.2in]{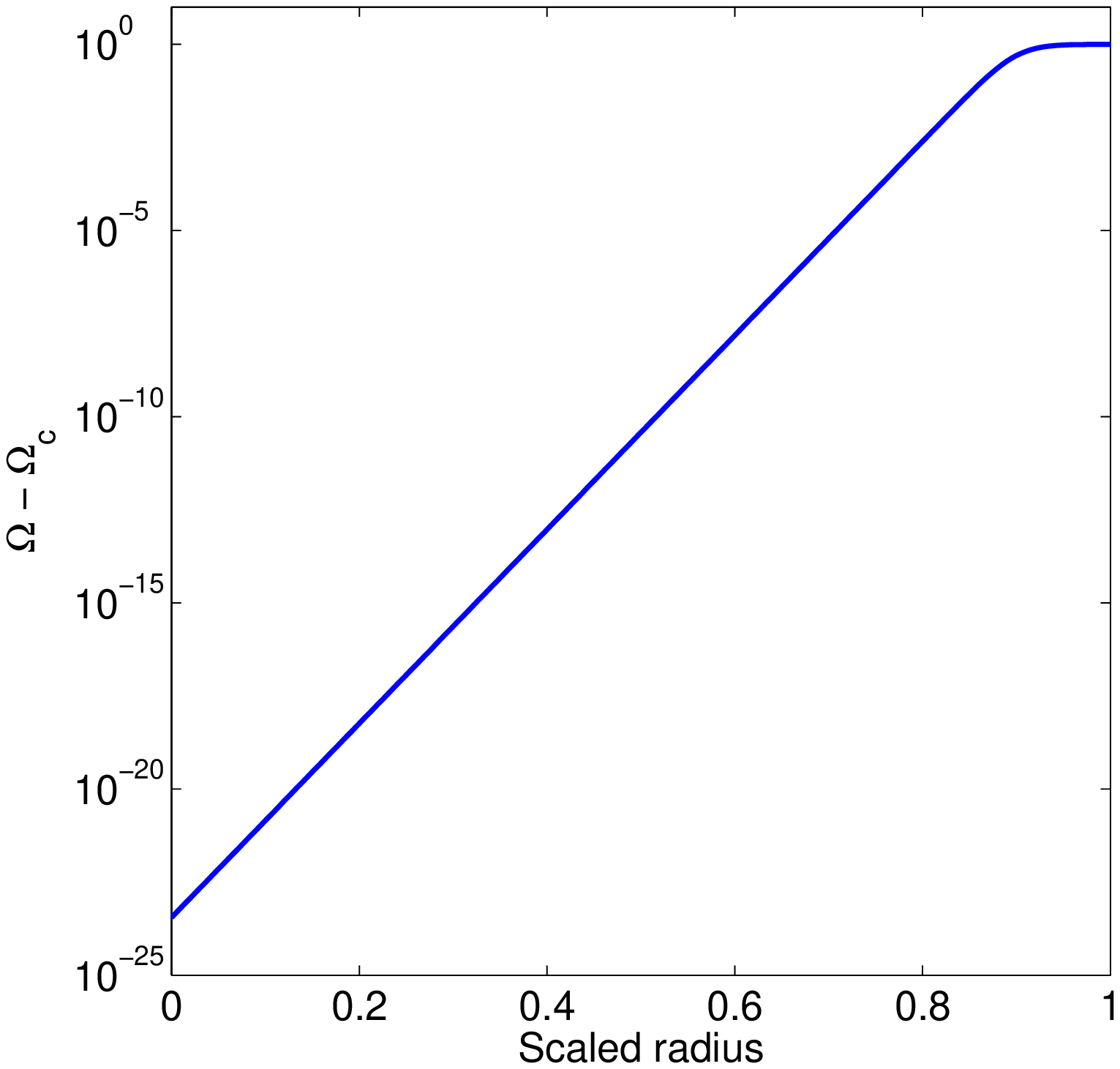} }
\caption{Angular velocity corresponding to current streamlines shown
in figure (\ref{odfig: current_density}). Parameters in equation
(\ref{flow}) are $r_{mp}/R = 0.9$, $c_{mp} = 50$. Surface value of
$\Omega_P + \Omega_{out}$ is scaled to be near unity. }
\label{specify_velo}
\end{figure}

\begin{figure}
\centerline{\includegraphics[width=2.2in]{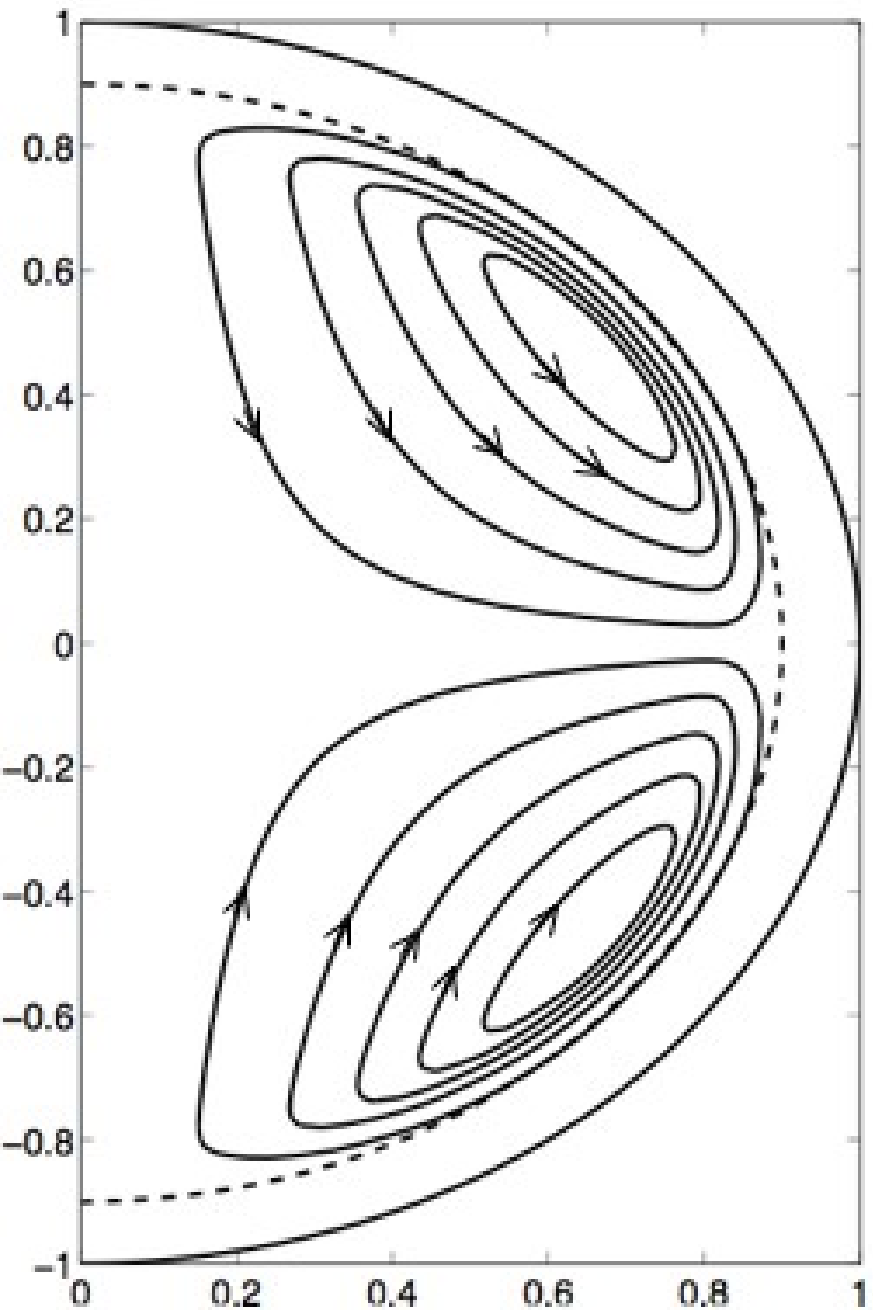} }
\caption{Current stream lines arising from a toy model. A vacuum
dipole field interacts with a simple cylindrical flow consisting of
a nearly uniformly rotating core and a more rapidly rotating
envelope. See text and figure (\ref{specify_velo}) for details. In
the outer envelope, the streamlines are close to lines of constant
$r$. Their spacing is inversely proportional to $jr\sin{\theta}$.
Ohmic dissipation per unit volume, $\propto\sigma U_\phi^2 B_p^2$,
is maximal near the dash line which marks to core-envelope boundary.
Our model only makes sense in regions where the magnetic Reynolds
number based on the convective velocity is small.} \label{odfig:
current_density}
\end{figure}

We calculate the current distribution arising from a vacuum axial
dipole field and a simple zonal flow. The angular velocity,
$\Omega$, is taken to be of the form
\begin{equation}
\Omega = \Omega_P + \frac{\Omega_{out}} {1 + \exp \left[c_{mp}
(r_{mp}-r) \right] } \, . \label{flow}
\end{equation}
Here $c_{mp}$ is the truncation factor and $r_{mp}$ is the
truncation radius. $\Omega(r) \rightarrow \Omega_P$ for $r < r_{mp}$
and $\Omega(r) \rightarrow \Omega_P + \Omega_{out}$ for $r >
r_{mp}$.\footnote{We plot $\Omega - \Omega_P$ as a function of
scaled radius in figure (\ref{specify_velo}).}

The current density arising from interaction of this flow with a
vacuum dipole magnetic field satisfies:
\begin{equation}
\bn \times\frac{\bj}{\sigma} =\bn \times\left\{ \left[
\left(\Omega-\Omega_P\right) \ez \times \br\right] \times \bb
\right\} = -\frac{2M}{3r^2}\frac{d\Omega}{dr}\frac{dP_2}{d\theta}
\ep \, ,
\end{equation}
where $M$ is the dipole moment.

Because $\bn\cdot\bj=0$, $\bj$ can be derived from a vector
potential $A$ such that
\begin{equation}
\bj = \bn \times (A \ep)\, ,\label{bj_simple}
\end{equation}
where
\begin{equation}
A = \frac{-rf(r)}{6} \frac{dP_2}{d\theta}\, .
\end{equation}
By direct computation
\begin{equation}
\bn\times\frac{\bj}{\sigma}
=\frac{1}{6r\sigma}\left[\frac{d^2}{dr^2}\left(r^2f\right) -6f
-\frac{d\ln\sigma}{dr}\frac{d}{dr} \left(r^2f\right)\right]
\frac{dP_2}{d\theta}\ep\, .
\end{equation}
So
\begin{equation}
\frac{d\Omega}{dr}=\frac{-r}{4\sigma M}
\left[\frac{d^2}{dr^2}\left(r^2f\right) - 6f
-\frac{d\ln\sigma}{dr}\frac{d}{dr}\left(r^2f\right)\right]\, .
\label{j_before_inte}
\end{equation}
At the planet's surface, the current density along the radial
direction goes to zero: $f(r) \rightarrow 0$; Near its center,
$j\propto r$: $df(r)/dr-f(r)/r \rightarrow 0$. We solve equation
(\ref{j_before_inte}) to obtain $f(r)$.

Since,
\begin{equation}
\bj \cdot \bn\left(A r \sin{\theta}\right) = 0\, ,
\end{equation}
the current density follows the contour lines of $A r \sin{\theta}$
and its magnitude satisfies $|\bj| = \left| \bn \left( Ar
\sin{\theta}\right) \right| / r \sin{\theta}$. Figure (\ref{odfig:
current_density}) displays the streamlines of current flow.




\end{document}